\documentclass[fleqn,usenatbib]{mnras}

\usepackage{newtxtext,newtxmath}

\usepackage[T1]{fontenc}

\DeclareRobustCommand{\VAN}[3]{#2}
\let\VANthebibliography\thebibliography
\def\thebibliography{\DeclareRobustCommand{\VAN}[3]{##3}\VANthebibliography}

\usepackage{graphicx}	
\usepackage{amsmath}	

\title[Carbonate not melilite]{Sakurai's Object revisited: new laboratory data for carbonates and melilites suggest the carrier of 6.9~\micron\ excess absorption is a carbonate}
\author[Bowey \& Hofmeister]{J. E. Bowey$^{1}$\thanks{E-mail:boweyj@cardiff.ac.uk}
and A. M. Hofmeister$^{2}$\\
$^{1}$School of Physics and Astronomy, Cardiff University, Queen's Buildings, The Parade, Cardiff, CF24 3AA, UK.\\
$^{2}$Department of Earth and Planetary Sciences, Washington University, 1
Brookings Drive, St. Louis MO 63130, USA.\\
}

\date{Accepted XXX. Received YYY; in original form ZZZ}

\pubyear{2021}

\begin{document}
\label{firstpage}
\pagerange{\pageref{firstpage}--\pageref{lastpage}}
\maketitle

\begin{abstract}
We present new room-temperature 1100--1800 cm$^{-1}$ spectra of
melilite silicates and 600--2000cm$^{-1}$ spectra of three randomly
orientated fine-grained carbonates to determine the possible
carrier(s) of a 6.9~\micron\ absorption feature observed in a variety
of dense astronomical environments including young stellar objects and
molecular clouds. We focus on the low-mass post-AGB star Sakurai's
Object which has been forming substantial quantities of carbonaceous
dust since an eruptive event in the 1990s. Large melilite grains
cannot be responsible for the 6.9~\micron\ absorption feature because
the similarly-shaped feature in the laboratory spectrum was produced
by very low (0.1 per cent by mass) carbonate contamination which was
not detected at other wavelengths. Due to the high bandstrength of the
6.9~\micron\ feature in carbonates, we conclude that carbonates carry
the astronomical 6.9~\micron\ feature.  Replacement of melilite with
carbonates in models of Sakurai's object improves fits to the
6--7~\micron\ Spitzer spectra without significantly altering other
conclusions of Bowey's previous models except that there is no link
between the feature and the abundance of melilite in meteorites. With
magnesite (MgCO$_3$), the abundance of 25-\micron-sized SiC grains is
increased by 10--50 per cent and better constrained. The mass of
carbonate dust is similar to the mass of PAH dust. Existing
experiments suggest carbonates are stable below 700~K, however it is
difficult to ascertain the applicability of these experiments to
astronomical environments and more studies are required.

\end{abstract}



\defcitealias{Evans2020}{Ev20}
\defcitealias{Bowey2021}{Bo21}
\defcitealias{BH2005}{BH05}

\section{Introduction}

A 0.5-\micron-wide 6.9~\micron\ absorption feature is of interest in a wide
astronomical context because the feature is seen in young stellar
objects and molecular clouds which have a complex inventory of dust
features due to ices, carbonaceous materials and silicates.  The
feature also seems to underly narrower PAH absorption in 2005--2008 Spitzer
spectra of dust obscuring the carbon-rich born again post-AGB star
known as Sakurai's Object (V4334 Sgr) \citep[][hereinafter
  \citetalias{Bowey2021}]{Bowey2021}.

Due to the chemical complexity of YSOs and molecular clouds various
carriers have been considered including carbonates although
\citet{Keane2001} ruled them out because the laboratory features were
too broad. Candidates also include mixtures including methanol ice or
another saturated hydrocarbon \citep{Tielens1984} and mixtures of
polycyclic aromatic hydrocarbons
\citep[PAHs, ][]{Mattioda2020}. \citet[][hereinafter
  \citetalias{BH2005}]{BH2005} noticed that the band seemed to occur
in environments with very deep 10~\micron\ silicate absorption bands
and found that very high column densities and/or relatively-large
(10~\micron-sized) crystalline melilite (silicate) grains might be
responsible.

Observational constraints were placed on putative band carrier(s) in a
large sample of YSOs and molecular clouds
\citep{Boogert2008,Boogert2011}. \citeauthor{Keane2001} and
\citeauthor{Boogert2008} consider the band to have components centred
at 6.75 (FWHM$\sim 0.23$~\micron) and 6.95 (FWHM$\sim0.30$~\micron)
finding the 6.95~\micron\ peak to exist in all environments while the
relative 6.75~\micron\ band strength decreases significantly in some
YSOs and the changes seem to correlate with the H$_2$O ice
abundance. \citet{Boogert2008} tentatively associated the NH$_4^+$ ion
in UV irradiated ice mixtures with the 6.9~\micron\ feature.



In contrast to these interpretations, carbonates are responsible for
6.9-\micron\ absorption in laboratory spectra of hydrated
interplanetary dust particles (IDPs) \citep[e.g.][]{SW1985} which were
assumed to originate from planetesimal and comet
environments. \citet{Lisse2007} included carbonates (magnesite
(MgCO$_3$) and siderite (FeCO$_3$)) in models of freshly produced dust
from the deep impact experiment on comet Temple 1 with the caveat that
they might be difficult to form in the absence of liquid
water. However, apparent difficulties in carbonate formation did not
prevent the identification of a far-infrared a band at $\sim
90$~\micron\ in the spectrum of PN NGC~6302 with calcite (CaCO$_3$)
\citep[e.g][]{Kemper2002} but its temperature has to be below 45~K to
explain the absence of a 44~\micron\ band \citep{Posch2007}. A
90~\micron\ emission band in YSOs has also been compared with
carbonates\citep[e.g][]{2002A&A...395L..29C,Chiavassa2005}.
\citet{Lisse2007} described the FIR identifications as problematic due
to the weakness of the laboratory features and possible confusion with
silicate emission bands.

A similar 6.9~\micron\ feature with a companion at 6.3~\micron\, was
identified in newly-formed ice-free dust surrounding the currently
carbon-rich photosphere of the low mass post-AGB star known as
Sakurai's Object (V4334) \citep[][hereinafter
  \citetalias{Evans2020}]{Evans2020}. Once the photosphere was
obscured, CN, C$_2$ and CO were detected in its atmosphere
\citep{Eyres1998}. Subsequent infrared observations revealed HCN and
C$_2$H$_2$ with $^{12}$C to $^{13}$C isotope ratios which are
consistently lower than typical Solar System values
\citepalias{Evans2020}. The star is surrounded by a compact
$30\times40$ milliarcsec opaque dusty disc or torus inclined at
75~$\deg$ to the plane of the sky \citep{Chesneau2009}. It should be
easy to decipher the carrier(s) of the 6.9-\micron\ features there
since the new dust is theoretically carbon-based.  However, beyond the
torus the object is surrounded by a faint circular planetary nebula
(PN) with a radius of 32~\arcsec which is aligned with the inner torus
and there is foreground silicate absorption \citep{Evans2002}.


\citetalias{Bowey2021} modelled the 6--7-\micron\ bands with a
combination of 30~nm-sized PAHs and 20~\micron-sized SiC and
10-\micron-sized melilite grains. Melilites (oxygen-rich) would not
normally be expected in a carbon-rich environment. However, fits to
the 8--13.3~\micron-range required additional absorption due to
submicron-sized \emph{interstellar} silicates, termed astrosilicates,
and represented by dust towards Cyg~OB2~no.~12. Since the increase in
the melilite feature seemed to occur with a decrease in the silicate
absorption feature, Bowey proposed that the melilites might be
converted from smaller silicate grains in the ancient PN.

Like others, \citetalias{Bowey2021} used the absence of a formation
mechanism to rule carbonates out of consideration in the H$_2$O-free
environment of Sakurai's Object. However, she was unaware that calcium
carbonate has been formed under laboratory conditions by exposing
amorphous CaSiO$_3$ (a glass of wollastonite composition) with gaseous
CO$_2$ at pressures > 6 Bar \citep{Day2013}.

A weakness in \citepalias{BH2005} and \citetalias{Bowey2021}
association of melilite with the astronomical 6.9~\micron\ bands is
its basis on a single laboratory spectrum and the absence of good
laboratory spectra for carbonates in this wavelength range. Therefore
we present experimental methods in Section~\ref{sec:expt}, new
laboratory spectra of carbonates in Section~\ref{sec:carbonates} and
more melilites in Sections~\ref{sec:fmelilites} and
\ref{sec:melilites}. Sakurai's Object is modelled with carbonates in
Section~\ref{sec:sakurai} and revised column densities, torus masses
and mass-increase rates are given in
Section~\ref{sec:massdensity}. Mechanisms for astronomical carbonate
formation are discussed in Section~\ref{sec:carbform} and the paper is
summarised in Section~\ref{sec:conclusion}.

\section{Experimental Methods}
\label{sec:expt}
Electron microprobe analysis of the new melilite samples was performed
at Washington University using a JEOL-733 equipped with Advance
Microbeam automation.  The accelerating voltage was 15 kV, beamcurrent
was 25 nA, and beam diameter was 10~$\mu$m. X-ray matrix corrections
were based on a modified \citet{Arm88} CITZAF routine. Silicates and
oxides were used as primary standards.

IR absorption spectra were acquired using an evacuated Bomem DA 3.02
Fourier transform spectrometer with an accuracy of $\sim 0.01$
cm$^{-1}$ and an SiC source. 1500 scans at 2-cm$^{-1}$-resolution were
obtained with a HgCdTe detector and a KBr beam splitter.

Powders were made by hand-grinding mineral samples to submicron ($\sim
0.1$--1-\micron) sizes under alcohol with a pestle in a ceramic mortar
and drying them for five minutes. To avoid degrading the crystal
structure grinding and drying were limited to 10 and 5 minute
intervals, respectively. Grain size was estimated by the feel of the
powder under the pestle; if sizes were too large the grinding
procedure was repeated.

Four powder spectra of known and unknown sample thickness were
gathered for each mineral. Powders were hand-compressed between two
25~mm-diameter 4~mm-thick KBr discs to make optically-thin films; a
reference spectrum of the KBr discs was obtained before applying the
sample. Discs were cleaned with lens tissues between runs but some
sample can remain in microscopic surface scratches so the discs were
reground with 3M polishing plastic with diamond coatings of 9, 5, and
1~\micron\ before taking measurements for a different mineral. A
single pair of discs was used for the Carrara marble, magnesite and
dolomite samples. Melilite spectra were measured with a different set
of discs.

Films of known thickness were measured by compressing powder
within apertures between the KBr discs. Apertures were made in the
$\sim 1.9$~\micron-thick microphone foils measured by \citep{BKS2020}
and with with 6~\micron-thick tin gaskets used for IR cells (these were
measured with a micrometer). The very thin films required to obtain
unsaturated carbonate peaks were obtained by gently squeezing the
discs together and rotating them to make a uniform film. Spectra of
each sample were taken in the order microphone foil, repeat after
squeezing, 6~\micron\ gasket, then take gasket out and run the very
thin film. The thinnest calcite film ($\sim 0.04$~\micron) is of grains
remaining in the scratched KBr discs after cleaning with a lens paper.

Melilite cleavage flakes (chips) were set on an aperture. The
unobstructed aperture was used for the reference spectrum. Thicknesses
were measured with a micrometer.

\subsection{Determination of band strengths and mass absorption coefficients}
Laboratory absorbance
\begin{equation}\label{eq:abar}
a=-\log_{10}\frac{I_t}{I_*}=A \times d,
\end{equation}
where $I_t$ is the intensity of the beam transmitted through the
sample and its holder and $I_*$ is the intensity of the beam
transmitted through the empty holder. Laboratory absorbance is
equivalent to the absorption coefficient, $A$, times the film
thickness $d$ in the spectroscopic, chemical and mineralogical
literature. However, astronomers use a natural log absorption
coefficient (or optical depth, $\tau$) units, so our data are plotted
as
\begin{equation}
\label{eq:tau}
  \tau (\mu m^{-1})=\frac{A}{d}\times2.3026,
\end{equation} where the factor of 2.3026 is derived from the change of base formula.


We present our data as natural log absorption coefficients because
this represents attenuation and path length. Mass extinction
coefficients, $\kappa$ (cm$^2$g$^{-1}$), can be calculated using the
relation
\begin{equation}
\label{eq:mabs}  
  \kappa=10^4\tau/\rho
\end{equation}
where $\rho$ is the density of the mineral in units of gcm$^{-3}$.

Carbonate spectra of nominally $\sim 0.1$-\micron-thick films were
scaled to the height\footnote{Peak areas were not used because
reflections contribute to the LO modes and asymmetry in carbonate
spectra. For weak modes the LO-TO splitting is low.} of the $\sim
712$~cm$^{-1}$ or 14.0~\micron\ E$\perp$c peak because it was
unsaturated in all the films and originates from the same orientation
as the dominant 1450-cm$^{-1}$ 6.9-\micron\ peak. ELF routines in the
STARLINK DIPSO software package were used to subtract local baselines
and to fit Gaussian peaks. The microphone foils were used as the
primary thickness gauge and the 6~\micron\ films for a consistency
check. The thickness of the $\sim$0.04-\micron\ film (powder remaining
in a scratch on the KBr disc after cleaning) was estimated by matching
712~cm$^{-1}$ and 876~cm$^{-1}$ bands to those in the
0.11~\micron\ calcite film. Polynomial baselines were subtracted from
the entire spectral range before publication in
Figure~\ref{fig:carbonate}, an unsubtracted spectrum of calcite is
shown in Figure~\ref{fig:contaminatedmelilite}. Fringe removal was
unnecessary.



Local baselines and Gaussian fits to peaks in the powder and chip
spectra of melilites Ak$_{70}$ and Ak$_{13}$ were obtained with ELF
routines. Ak$_{70}$ overtones were scaled to the height of the
1470cm$^{-1}$ (6.80~\micron) peak.  To ensure the absence of carbonate
and other contaminants, the spectrum of, Ak$_{13}$, was obtained after
soaking it in muriatic acid (concentrated HCl) for a few days, drying
it, and subtracting the spectrum of an unidentified fluffy residue
from the resulting crystals. The baseline subtraction of Ak$_{70}$ and
the subtraction of the spectrum of the residue from the spectrum of
Ak$_{13}$ is described in Appendix~\ref{app:base}.


\section{Carbonate Spectra}
\label{sec:carbonates}
\begin{table*}
\begin{minipage}{\linewidth}
  \caption{Carbonate samples, peaks and estimated band strengths for $\sim 0.1$~\micron\ powder films. \label{tab:carbonate}}
\begin{tabular}{llllll}
\hline
Sample &Formula&Locality\footnote{See  \citep{Merriman2018} for sample details.\label{foot:mer}}&Peaks&$\rho^{\ref{foot:mer}}$&$\kappa_{6.9}$\\
&&&\micron&gcm$^{-3}$&cm$^2$g$^{-1}$\\
\hline
Calcite& CaCO$_3$&Carrara, Italy&5.57, {\bf 6.97}, 11.42, 14.03  &2.72&49000\footnote{for 0.11-\micron\ film; value for very thin film is $\approx 64000$}\\
Dolomite&CaMg(CO$_3$)$_{2}$&Eugul, Navarre, Spain&5.50, {\bf 6.87}, 11.33, 13.70&2.86&33000\\
Magnesite\footnote{composition from the RRUFF Database https://rruff.info/ \citep{RRUFF}}&MgCO$_3$&Brumado, Bahia, Brazil&5.46, {\bf 6.87}, 11.25, 13.37&2.98&27000\\
\hline
\end{tabular}
\end{minipage}
\end{table*}
\begin{figure}
\includegraphics[bb=125 285 452 650,width=\linewidth,clip=]{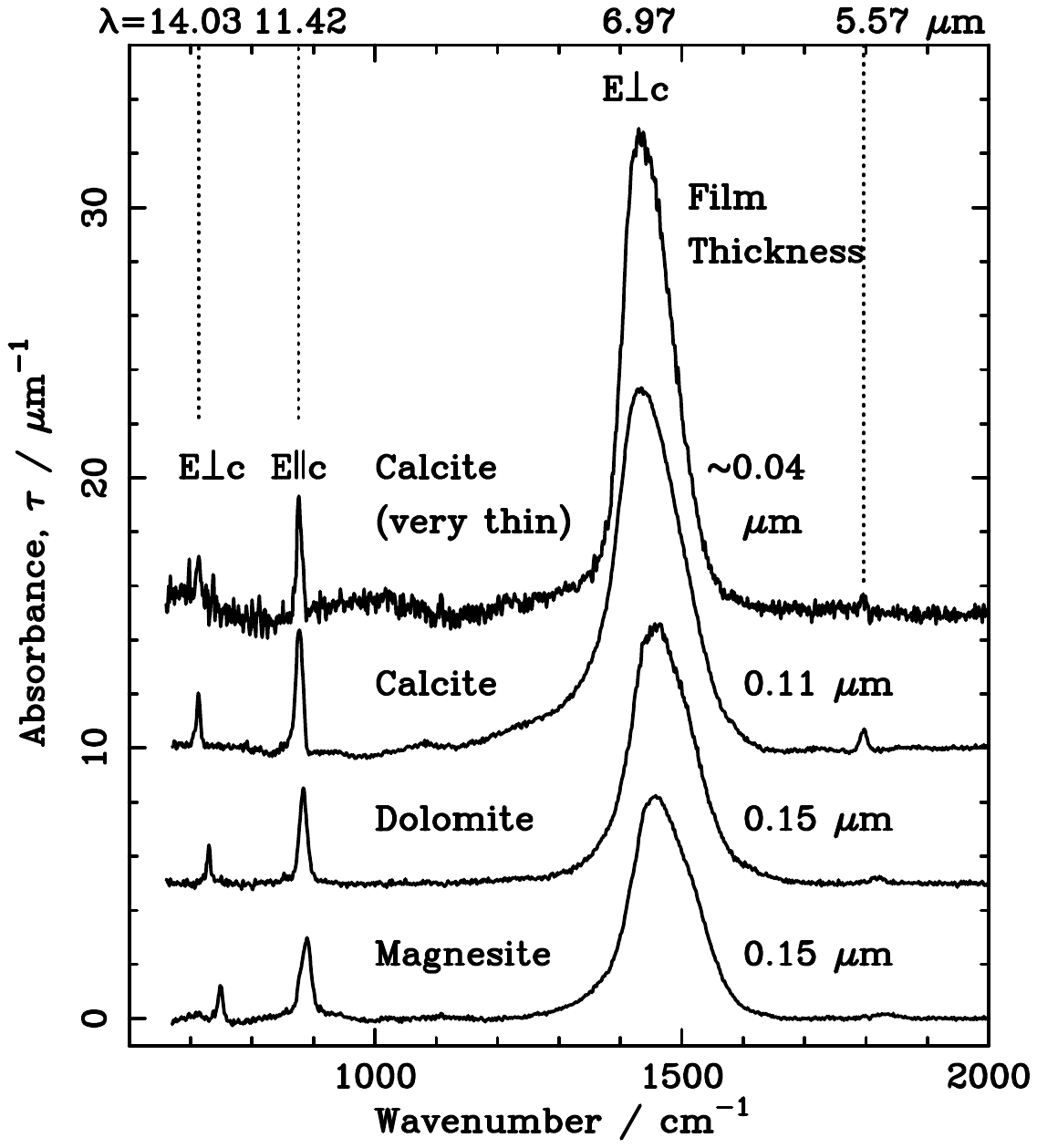}
\caption{Calibrated carbonate spectra. From bottom to top offsets in y
  are 0, 5, 10 and 15. Differences in band strengths are intrinsic to
  the minerals.}
    \label{fig:carbonate}
\end{figure}

We selected marble specimens ( e.g. Carrara marble) which are fine
grained with randomly orientated crystals, because most infrared
carbonate bands depend strongly on crystal orientation \citep[see
  review by][]{White1974}. Spectra of calcite, CaCO$_3$ dolomite
(Mg$_{0.5}$Fe$_{0.5}$CO$_3$) and magnesite (MgCO$_3$) are presented in
Figure~\ref{fig:carbonate} and the peaks listed in
Table~\ref{tab:carbonate}. Siderite, FeCO$_3$, was not measured
because its peak occurs longward of the astronomical band at $\sim
7.03$~\micron\ \citep{White1974}. Carbonate spectral features in the
600--2000~cm$^{-1}$-range are due to stretches within the CO$_3^{2-}$
anion and wavelength shifts due to changes in cation are relatively
small at $\sim \pm 0.1~\micron$ for the dominant
6.9-\micron\ feature. In calcite, 1434~cm$^{-1}$ (6.97~\micron) and
narrow 713~cm$^{-1}$ (14.0~\micron) peaks are sensitive to E $\perp$
c, whilst the 876~cm$^{-1}$ (11.4~\micron) is sensitive to E
$\parallel$ c. The weak 1796~cm$^{-1}$ (5.57~\micron)-band is a
multiphonon mode without an orientational dependence. Band strength
estimates are high, $\kappa_{6.9}\sim 27000$--49000~cm$^2$g$^{-1}$ for
$\sim$ 0.1~\micron\ films; differences in band strengths are intrinsic
to the minerals.

\section{False melilite overtone spectra}
\label{sec:fmelilites}
The melilite group contains paired SiO$_4$ or SiO$_4$--AlO$_3$
tetrahedra with shared oxygen atoms; the main group ranges from \aa
kermanite Ca$_2$MgSi$_2$O$_7$ to gehlenite (Ca$_2$Al$_2$SiO$_7$);
intermediate melilites have varying combinations of Ca$^{2+}$, Na$^+$,
K$^+$, Mg$^{2+}$, Fe$^{2+}$ and Fe$^{3+}$. Unfortunately carbonate
contamination is almost inevitable due to the environments in which
melilites are formed. Carbonates and other contaminents are adequately
removed by examining the crystals before grinding. Hereinafter,
contaminated overtone spectra are distinguished by referring to them
as false overtones (fo), foAk$_{x}$, where $x$ represents the \aa
kermanite component of the contaminated overtone spectrum. Melilites
form from Ca-rich magmas and within thermally metamorphosed carbonate
rocks \citep{DHZ}. Synthetic samples are formed by melting
(foAk$_{100}$) or sintering (foAk$_0$) mixtures of CaCO$_3$,
Al(OH)$_3$, MgO and SiO$_2$ at 1300$^\circ$C \citep{Charlu:1981}.

\begin{figure}
\includegraphics[bb=125 205 452 640,width=\linewidth,clip=]{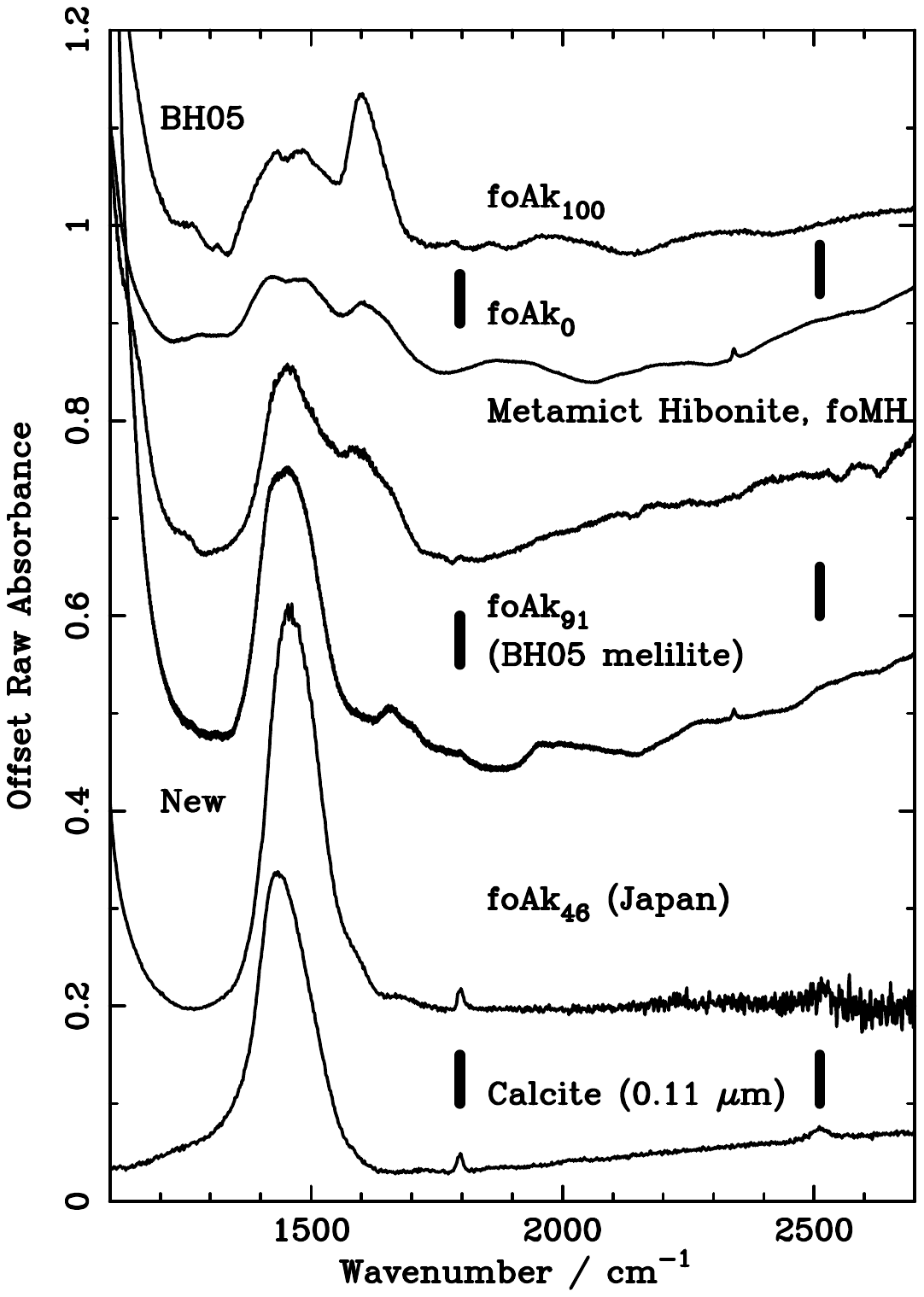}

\caption{Carbonate contamination in published (\citetalias{BH2005}) and new melilite (Japan) melilite spectra and calcium carbonate (0.11-\micron-thick Carrara Marble (CaCO$_3$)) before baseline subtraction. Thick bars indicate the expected position of small carbonate bands.}
    \label{fig:contaminatedmelilite}
\end{figure}

Peaks in the 5--8-\micron\ spectra measured by \citetalias{BH2005} are
compared with the spectrum of the 0.11-\micron-thick calcite film and
an additional melilite sample, foAk$_{46}$, from Japan
(Figure~\ref{fig:contaminatedmelilite}) before baseline subtraction. In
addition to the prominent 1434~cm$^{-1}$ band, narrow calcite peaks
are clearly visible in the Japan sample at 1796~cm$^{-1}$ and
2511~cm$^{-1}$. Its electron microprobed composition is 94~per~cent melilite
(Ca$_{1.94}$Na$_{0.07}$Fe$_{0.08}$Mg$_{0.33}$Al$_{0.60}$(Al$_{0.54}$Si$_{1.46}$)O$_7$),
5 per cent calcite and 1 per cent iron sulphide. Since silicate
overtones for most anhydrous silicates had
\citepalias[$\kappa_{pk}\sim 30$--90;][]{BH2005} it is probable
that the higher band strengths of crystalline melilites
$\kappa_{pk}\sim 130$--220 were enhanced by the carbonate impurity.

\citetalias{BH2005} and \citetalias{Bowey2021} found foMH and
foAk$_{91}$ matched astronomical 6.9~\micron\ features
well. Unfortunately, both spectra have a tiny inflections at
1796~cm$^{-1}$ which are probably the narrow calcite peak. The
synthetic end-member \aa kermanite (Ak$_{100}$) also had a weak peak
at 1796~cm$^{-1}$ but there is no hint of this in gehlenite
(Ak$_{0}$). Other calcite peaks at 876~cm$^{-1}$ and 712~cm$^{-1}$
were invisible due to saturation of the fundamental bands, which is
necessary to view the overtones.

In summary, the \citetalias{BH2005} overtone spectra with strong
6.9~\micron\ features are dominated by the carbonate
contamination. \citetalias{BH2005} spectra of materials without a
6.9~\micron\ band are not contaminated with carbonates.  This means
that \citetalias{BH2005} and \citetalias{Bowey2021} unknowlingly
fitted astronomical data with carbonates rather than melilites.  The
carbonate band is $\sim 1000$ times stronger than silicate overtones
and 5000 times the hibonite overtones, so we suspect the
contamination to have been $<1$ part in 1000 and otherwise
undetectable in the melilite spectra.


\section{True melilite overtone spectra}
\label{sec:melilites}
\begin{table}
\begin{minipage}{\linewidth}
  \caption{Spectral characteristics, composition and origins of powdered carbonate-free melilites with well-defined overtone spectra.}
  \label{tab:trumel}
  \begin{tabular}{llll}
\hline
Sample &Chemical Formula&Locality\footnote{ExM--Excalibur Mineral Co.}  &Peaks\\ 
&&&\micron\\ 
\hline
Ak$_{13}$&(Ca$_{2.02}$Na$_{0.07}$)(Fe$_{0.01}$Mg$_{0.15}$Al$_{0.83}$)&Crestmore, CA&6.42\\
&\hspace{2cm} --(Al$_{0.87}$Si$_{1.13}$)O$_7$&USA (Wards)&6.84\\
Ak$_{70}$\footnote{composition of 120~\micron\ chip}  &Ca$_{1.82}$(Fe$_{0.17}$Mg$_{0.22}$Al$_{0.62}$)&Mt Monzani&6.39\\
&\hspace{2cm} --(Al$_{0.30}$Si$_{1.70}$)O$_7$&Italy (ExM)&6.80\\
\hline
\end{tabular}
\end{minipage}
\end{table}

\begin{figure}
\includegraphics[bb=120 205 452 590,width=\linewidth,clip=]{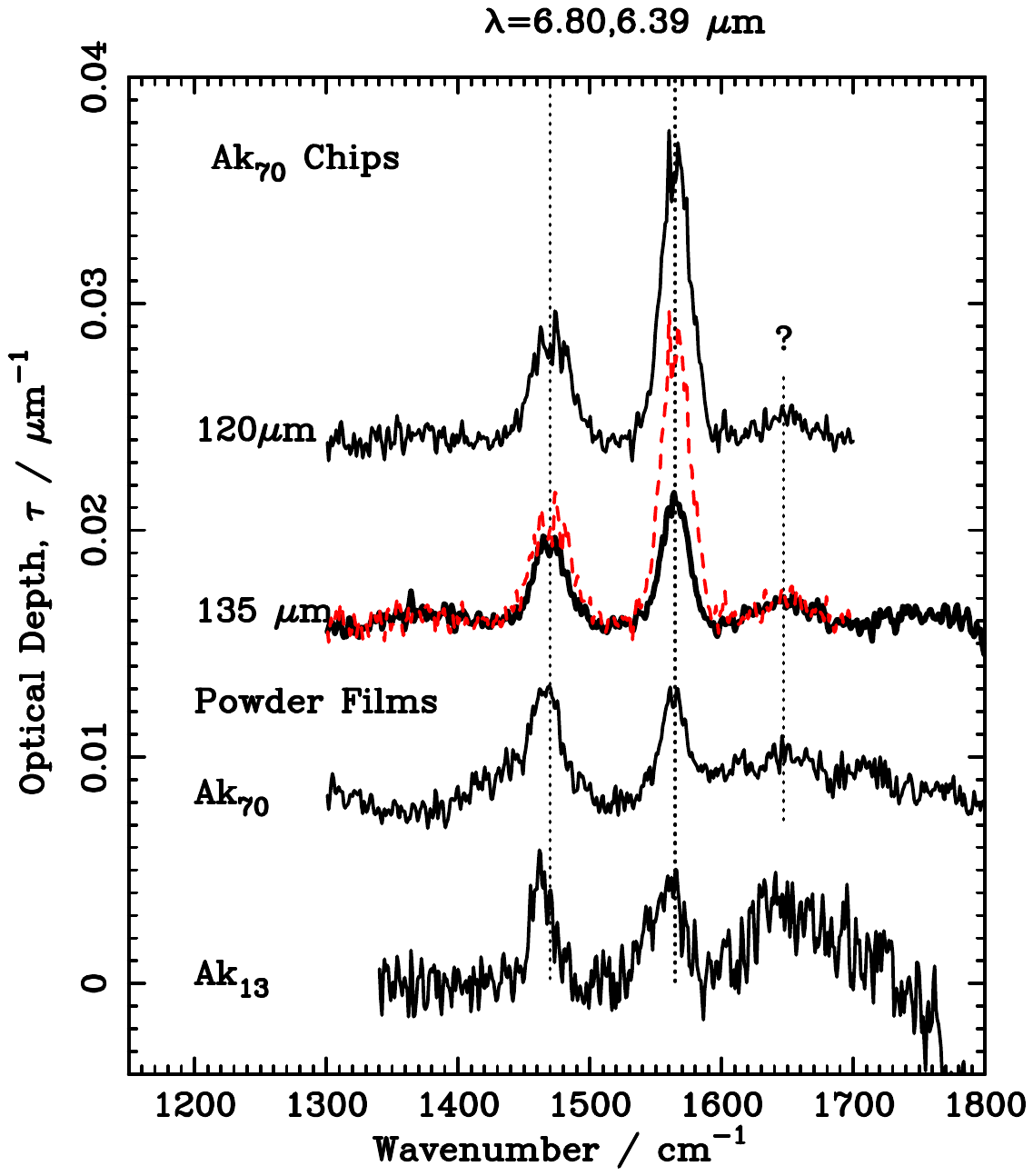}

\caption{True melilite overtones. The mineral chips show orientational effects. y-axis offsets bottom to top are: 0.000, 0.008, 0.016, 0.024}
    \label{fig:trumel}
\end{figure}

Spectra were obtained from 5--8~\micron\ of 7 additional melilites. However
the spectra were inconsistent and contained a range of broad bands
similar to those in Figure~\ref{fig:contaminatedmelilite}. Two samples
contained very different bands: an Italian melilite, Ak$_{70}$, and a
gehlenite from Crestmore, California, Ak$_{13}$. To ensure the absence
of carbonate and other contaminants, the spectrum of, Ak$_{13}$, was
obtained after soaking it in muriatic acid (concentrated HCl) for a
few days, drying it, and subtracting the spectrum of a fluffy
residue from the resulting crystals. The sample was so small that this
provides a consistency check on the shape of the Ak$_{70}$ overtone
spectrum, but is not suitable for data modelling.

The overtone spectra of Ak$_{70}$ chips and powder films
(Figure~\ref{fig:trumel} and Table~\ref{tab:trumel} have two bands
centred at 1470 (FWHM$\sim 29$~cm$^{-1}$, and 1565~cm$^{-1}$
(FWHM$\sim 24$~cm$^{-1}$).  The powder peaks are of similar strength
($\kappa_{6.80} \sim 15$~cm$^2$g$^{-1}$), but the relative strengths
from the chips vary: equal in the 135-\micron-thick sample, but in the
120-\micron\ sample the 1565~cm$^{-1}$ peak is twice as strong as the
1470~cm$^{-1}$ peak. We suspect a dependence on crystal orientation
with 1565~cm$^{-1}$ E$\parallel$ c and the other parallel to the
a-axis. The extremely weak third peak centred at 1650~cm$^{-1}$
(FWHM$\approx 40$) required for baseline determination might be a
H$_2$O bend from surface sorbed water from the air, an overtone or a
combination of the two\footnote{The four or five evenly-spaced
$\tau\la 0.001$-\micron$^{-1}$ peaks underlying features in the 120
and 130~\micron\ chips might be due to fringing caused by the aperture
or crystal-thickness.}.







\begin{figure}
  \includegraphics[bb=100 85 495
    300,width=\linewidth,clip=]{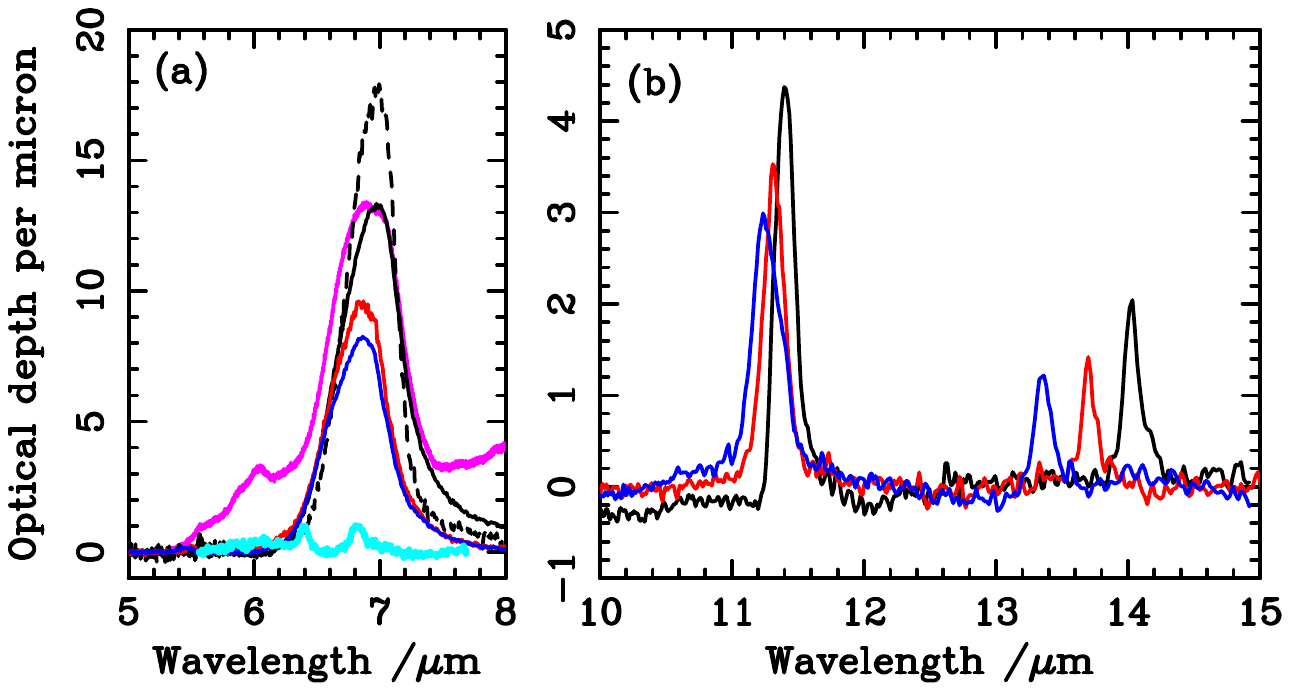}
\caption{(a) Wavelength comparison of 0.1--0.15-\micron-thick calcite
  (black solid curve), dolomite (red), magnesite (blue) with the \citetalias{BH2005} contaminated melilite fo$Ak_{91}$ (magenta) and the true
  melilite overtones, Ak$_{70}$ (cyan). The black dashed curve denotes the spectrum of the very thin$\sim$0.04~\micron\ calcite film. (b) 0.11-0.15~\micron-thick carbonate peaks between 10 and 15~\micron.}
\label{fig:wcarbmelcomp}
\end{figure}

The carbonate spectra are compared with the \citetalias{BH2005}
melilite and true melilite overtones on a wavelength scale in
Figure~\ref{fig:wcarbmelcomp}.  Although the true melilite overtones
might contribute to astronomical 6.9~\micron\ features, their relative
weakness makes it unlikely that they would ever be observed.

\section{Carbonate models of Sakurai's Object}

True melilite overtones have not been included in these new models of
Sakurai's object because they are too narrow and very weak in
comparison to the carbonate and PAH bands.

\label{sec:sakurai}
\subsection{Models}
\label{sec:sl2fits}

\begin{table*}
\centering
\caption{Laboratory data and band asignments of the C-rich (PAH, bSiC, nSiC and magnesite) and O-rich (foAk$_{91}$ and astrosilicate) dust components used in the models of Sakurai's Object; mass absorption coefficients are given at the peak wavelengths marked in bold type; see \citetalias{Bowey2021} for their derivation. Values for the false melilite, foAk$_{91}$ used by \citetalias{Bowey2021} are included for comparison.}
\label{tab:lab}
\begin{minipage}{0.75\linewidth}
\begin{tabular}{llllllll}
\hline
Sample&Strongest Band&Approximate&Size\footnote{Representative grain length assuming approximately cubic geometry.}&$\rho$&$\kappa_{pk}$&$m_g$\footnote{Representative mass of single grain, assuming volume=(size)$^3$ }&Ref\footnote{Reference for spectrum 1--\citealt{Carpentier2012}; 2--\citealt{Hof2009};  3--\citealt{Speck2005}}\\
&&Assignment&&gcm$^{-3}$&$10^2$cm$^2$g$^{-1}$&g&\\
\hline
PAH&{\bf 6.3}&Arom. C=C&53$_{30}^{70}$~nm&$0.39^{0.2}_{1.8}$&$600_{130}^{1200}$\footnote{Best estimates. The superscripts and subscripts indicate the range depending on the effective mass
  density of the soot sample, the value of 1200 pertains to a grain of 30nm and bulk mass density}&$5.8\times 10^{-17}$&1\\
bSiC&\bf{6.6}&Overtones&25~\micron&3.2&2.4&5.0$\times 10^{-8}$&2\\
nSiC&{\bf 12.3}&Si--C stretch&3~nm&3.2&$15000$&$8.6\times10^{-20}$&2, 3\\
magnesite&6.87&&0.15~\micron&2.98&270&$1.0\times10^{-14}$&\\


astrosilicate\footnote{Represented by the line of sight to Cyg OB2 no.12; $\rho$ is an estimate given by the mean of forsterite and enstatite mass densities; $\kappa_{pk}$ from \citet{BA2002}.}&{\bf 9.8}&Si--O stretch&0.3~\micron&3.3&26&$8.9\times 10 ^{-14}$&\\
\hline
\multicolumn{5}{l}{\citetalias{Bowey2021} false melilite}\\
foAk$_{91}$&6.1, {\bf 6.9}&Overtones&12~\micron&3.0&2.2&$5.2\times 10^{-9}$&\\
\hline
\end{tabular}
\end{minipage}
\end{table*}

As in \citetalias{Bowey2021}, the Spitzer 5.9--7.5~\micron\ flux
spectra were fitted with an obscured black body model with up to three
foreground absorption components. Parameters of the PAH and SiC (or bSiC,
where b means big to represent the 25-\micron-sized grains) and
magnesite components required for the models are listed in
Table~\ref{tab:lab}. The flux, $F_\nu$ is given by:
\begin{equation}
F_\nu=c_0B_\nu(T)\exp{(-\sum^3_{i=1} c_i\uptau_i(\lambda))},
\label{eq1}
\end{equation}
where $B_\nu(T)$ is the Planck function, and $\uptau_i$ is the shape
of the $i^{th}$ absorption feature, normalised to unity at the tallest
peak in the wavelength range of interest, and $c_0$ and $c_1$ to $c_3$
are the fitted scale factors. To determine $c_0$, the
continuum was matched to the feature at 6.0~\micron, i.e. the
foreground absorption was assumed zero at this wavelength.

In \citetalias{Bowey2021} the source flux, $F_{\nu*}$ for fitting the
8.4--13.3~\micron~spectra was given by,
\begin{equation}
F_{\nu*}\propto B_\nu(T_1)+B_\nu(T_2)c_{nSiC}\uptau_{nSiC}(\lambda)
\end{equation}
where T$_1$ and T$_2$ are the temperatures of the optically-thick and
optically-thin emission components and $c_{nSiC}$ is a scaling
constant for normalised emission due to nm-sized SiC grains,
$\uptau_{nSiC}(\lambda)$. This emitted continuum was then extinguished
by foreground astrosilicate dust to give the observed flux, F$_\nu$,
\begin{equation}
  F_\nu =c'_0F_{\nu*}\exp{(-c_{pah}\uptau_{pah}(\lambda)-c_{sil}\uptau_{sil}(\lambda))}
\label{eq:oldmod}
\end{equation}
where, scaling constant $c'_0$, T$_2$ and $c_{sil}$ were fitted by $\chi^2$-minimization and 
$c_{pah}$ was fixed to the value obtained from the
5.9--7.5-\micron\ fits. 

The effect of including the carbonate fraction modelled in the 5.9--7.5-\micron~fits was considered by including it as an additional absorption component.
\begin{equation}
  F_\nu =c'_0F_{\nu*}\exp{(-c_{pah}\uptau_{pah}(\lambda)-c_{sil}\uptau_{sil}(\lambda)-c_{car}\uptau_{car}(\lambda))}
\label{eq3}
\end{equation}
where $c_{car}$ was fixed to the value obtained from the
5.9--7.5-\micron\ fits.  Continua for merged 5--13.5~\micron\ spectra
were derived by setting $c_{pah}$, $c_{sil}$ and $c_{car}$, to zero
and extending the wavelength range.
\begin{figure}
\includegraphics[bb=125 160 440 400,width=\linewidth,clip=]{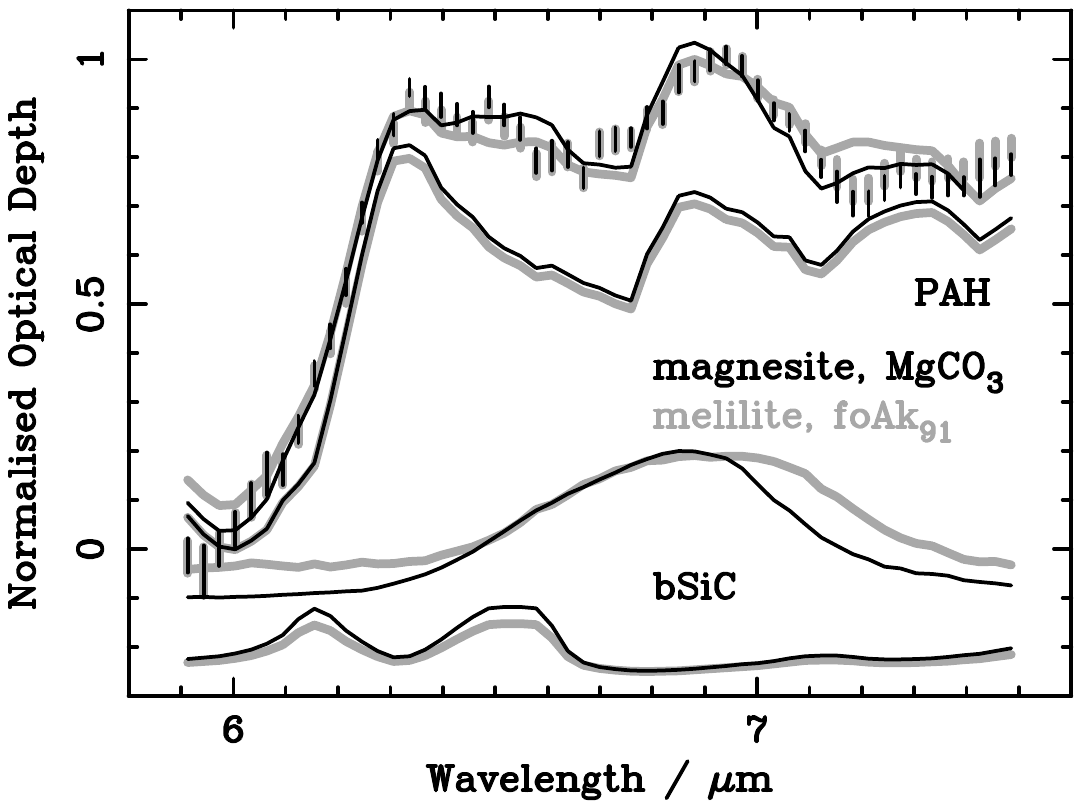}

\caption{Weighted mean of normalised optical depths obtained with
  foAk$_{91}$ melilite in \citetalias{Bowey2021} (thick grey error
  bars) and new magnesite (black error bars) with fits and indicated
  components - thick grey curves denote fits with melilite, black
  curves denote fits with magnesite. Profiles, fits and fitted
  components are very similar but the fitted bSiC component is
  slightly larger with magnesite.\label{fig:moptplot}}
\end{figure}

\subsection{Fits with carbonate change little}
\begin{table*}
\centering
\caption{Magnesite, MgCO$_3$, PAH and SiC fits to
  5.9--7.5~\micron~spectra of Sakurai's object. The 1 sigma confidence
  intervals $\sigma$, for $c_{pah}$ are 1--2 per cent; for $c_{mag}$
  they are normally 5--6 per cent and 8 per cent in
  Epoch~1. Quantities which vary by more than 10 per cent from those
  in \citetalias{Bowey2021} are indicated in bold.  Fit quality,
  $\chi^2_{\nu}$ and $\chi^2_{\nu2}$ denote three-component and
  two-component (PAH and melilite only) fits,
  respectively. $\Delta\chi^2_{\nu}$ is the difference between the
  qualities of the foAk$_{91}$ and the better magnesite
  fits. $\uptau_{6.3}$ and $\uptau_{6.9}$ are the measured optical
  depths of the 6.3 and 6.9~\micron\ peaks, respectively.}
\label{tab:sl2fits}
\begin{minipage}{\linewidth}
\begin{tabular}{lllllllllllll}
\hline
Ep. &MJD\footnote{Modified Julian Date (MJD) is used to identify the time of the observations. MJD is related to Julian Date (JD) by MJD = JD -- 2400000.5}&T/K  &$c_{pah}$&${c_{mag}}$&$c_{SiC}$&$\sigma$&$\chi^2_{\nu}$&$\chi^2_{\nu2}$&$\Delta\chi^2_{\nu}$&$\uptau_{6.3}$&$\uptau_{6.9}$&$c_{sil}$\\
\hline
1  &53475 &276 &0.080  &  0.033  &{\bf 0.013} & 32 & 0.56&0.62&0.34 &0.097&0.10&\\
2  &54225 &228 &0.12   &  0.057  &{\bf 0.016} & 20 & 0.64&0.86&0.32 &0.14 &0.15&0.072 \\
3  &54388 &226 &0.15   &  0.067  &{\bf 0.025} & 21 & 0.64&0.83&0.35 &0.18 &0.19&0.16  \\
4A &54577 &224 &0.19   &  0.065  &{\bf 0.028} & 28 & 0.52&0.74&0.25 &0.23 &0.23&0.14\\
4B &54586 &230 &0.21   &  {\bf 0.059}  &{\bf 0.029} & 14 & 0.49&0.78&0.030&0.23 &0.24&0.12\\
5  &54757 &228 &0.23   &  0.075  &{\bf 0.040} & 13 & 0.58&0.77&0.12 &0.26 &0.28&0.11\\
\\
WM&  &   &0.82   &  0.30   &{\bf 0.13}  &    &3.3&&1.0    &  0.94 &1.0\\
\hline
\end{tabular}
\end{minipage}
\end{table*}
\begin{figure*}
  \includegraphics[bb=23 40 505 755,height=0.8\linewidth,clip=,angle=-90]{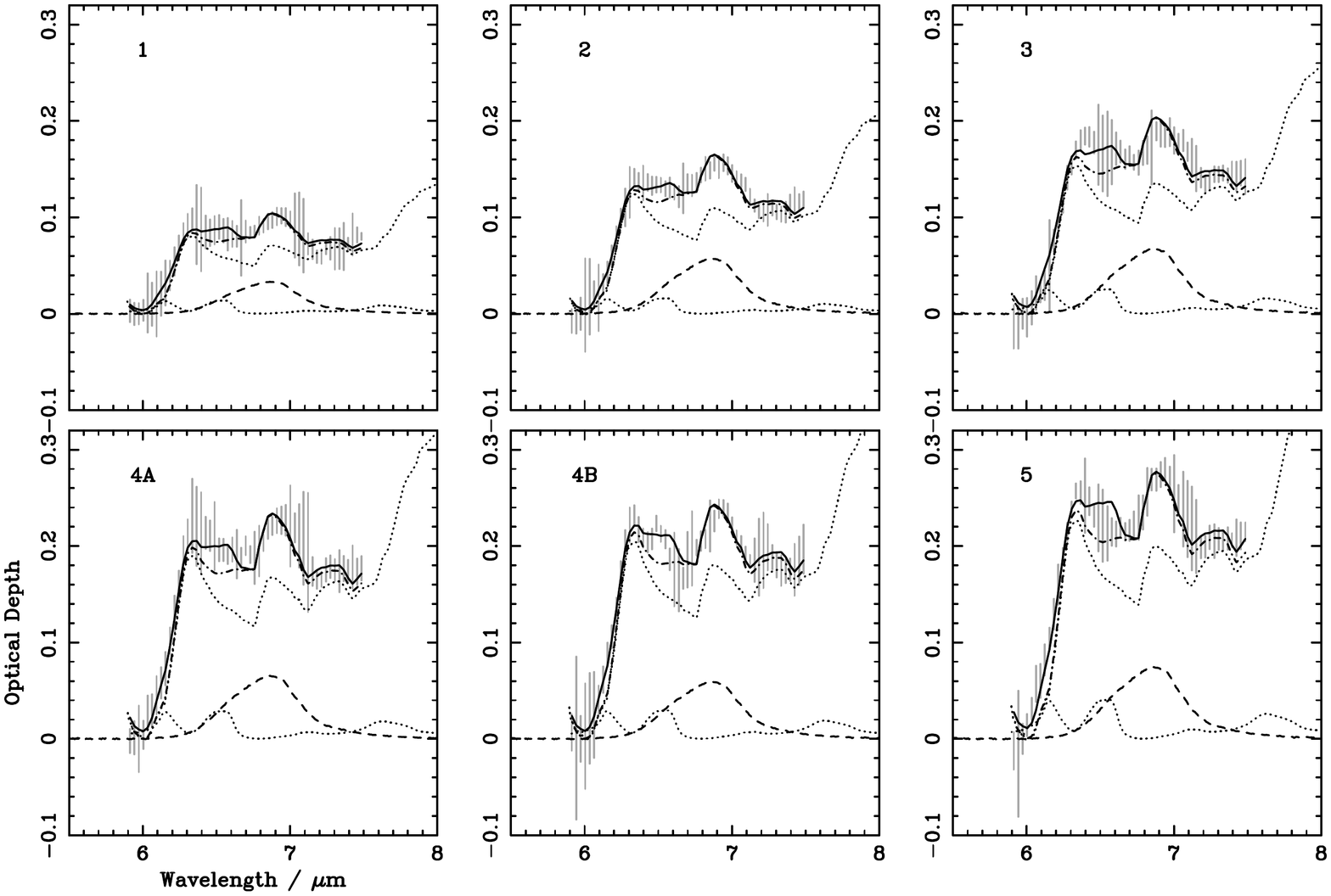}
\caption{Optical depth profiles of Sakurai's Object (error bars) for
  the 6--7.5-\micron\ range with fits (solid); fitted PAH (dotted
  -top), bSiC (dotted) and magnesite (dashed) components. Dot-dash
  curves indicate the effect of removing the bSiC component from the
  fits.\label{fig:optplot}}
\end{figure*}

All three carbonates provided good fits to the excess
6.9~\micron\ absorption in equation ~\ref{eq1}. Statistics of the
calcite models were similar to those of melilite (foAk$_{91}$) in
\citetalias{Bowey2021}, but fits with dolomite and magnesite were
slightly better due to a better match in peak wavelength.  Dolomite
and magnesite models were statistically indistinguishible from each
other ($\Delta \chi^2_\nu \pm$ 0.00--0.03) and variation in $c_{pah}$,
$c_{SiC}$ and $c_{mag}$ and $c_{dol}$ was $\sim \pm 0.01$. Since Mg is
estimated to be 12 times more abundant in grains than Ca
\citep{SW1996}, magnesite was chosen to represent the carbonate in
Sakurai's object. Differences in the fitted parameters are due mainly
to the nearly symmetrical shape of carbonate bands in comparison to
foAk$_{91}$ which has additional structure and subtantial opacity
beyond the 6.9-\micron\ band in foAk$_{91}$ (see Figure
~\ref{fig:wcarbmelcomp}).


\subsubsection{Virtually indistinguishable optical depth profiles}
The magnesite optical depth profiles and fit components for each epoch
are shown in Figure~\ref{fig:optplot} and listed in
Table~\ref{tab:sl2fits}.  Revised weighted mean optical depth spectrum
(black) and components are compared with the \citetalias{Bowey2021}
results (grey) in Figure~\ref{fig:moptplot}.  The revised and original
weighted means are virtually indistinguishable despite subtle differ-
ences in the fitted components.

\subsubsection{Higher and consistent $c_{SiC}$ estimates for different Epochs}
\begin{figure}
\includegraphics[bb=125 205 452 590,width=0.8\linewidth,clip=]{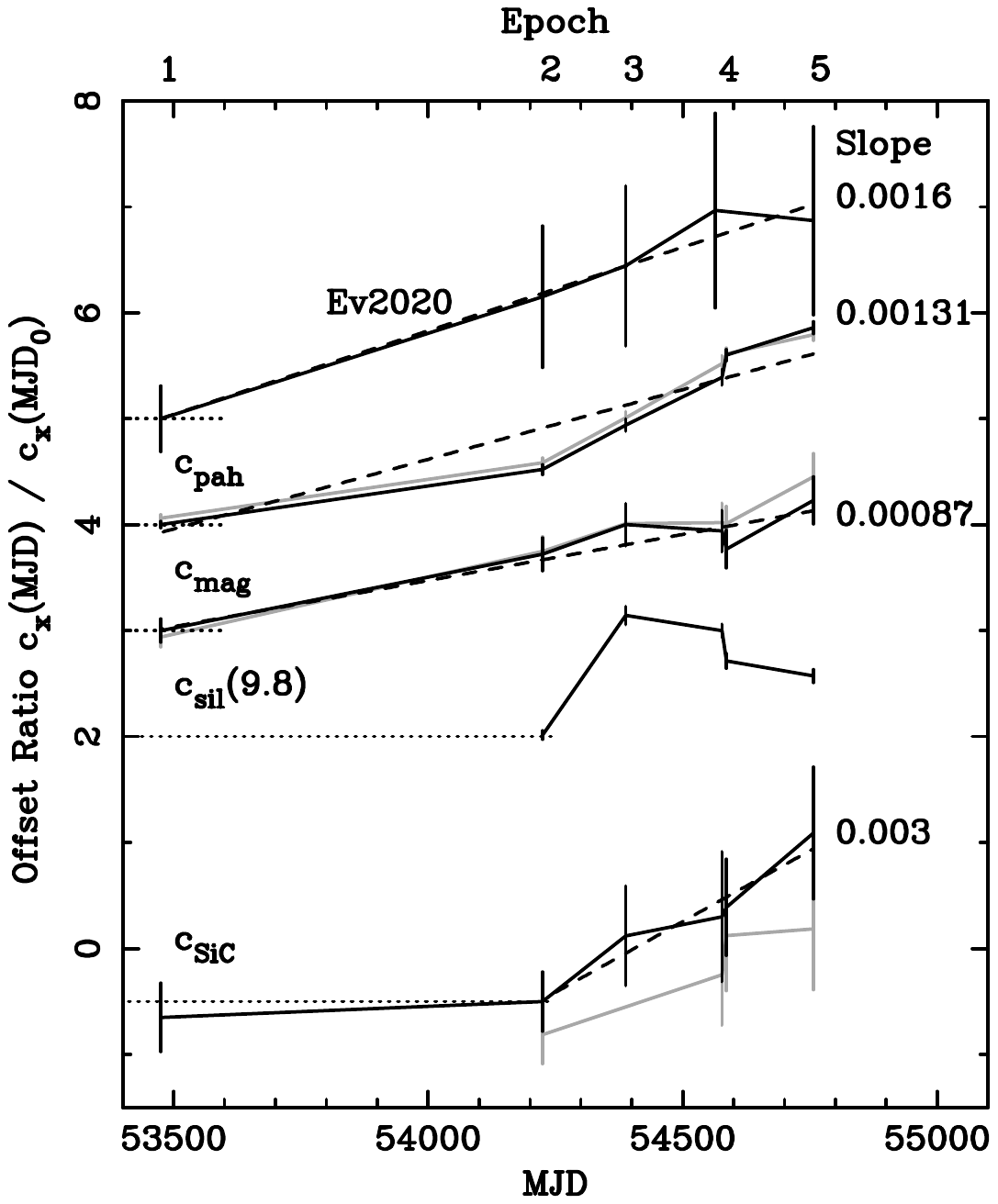}
\caption{Increases in dust mass during the Spitzer observing period
  (black) compared with those obtained with melilite by
  \citetalias{Bowey2021} (grey). Values obtained by
  \citetalias{Evans2020} are from the flux method. $c_{pah}$,
  $c_{mag}$ and $c_{SiC}$ are the fitted optical depths of PAHs and
  magnesite and bSiC, respectively. Melilite and magnesite indicators,
  except $c_{SiC}$, are ratioed to values obtained with magnesite at
  Epoch~1 (2005-04); $c_{SiC}$ is ratioed to its magnesite value at
  Epoch~2 (2007-05).  Offsets are indicated by dotted lines. Slopes in
  the figure are in optical depth per day, Uncertainties in the fitted
  slopes are 0.0005 \citepalias{Evans2020}, 0.00004 ($c_{pah}$),
  0.0001 ($c_{mag}$) and 0.001 ($c_{SiC}$) per day.\label{fig:dmi}}
\end{figure}

Ratioed PAH, magnesite and bSiC optical depths are plotted against
their modified julian dates (MJD) in Figure~\ref{fig:dmi}; values
obtained with foAk$_{91}$ are indicated in grey. For consistency with
~\citetalias{Bowey2021}, the fitted optical depths, except $c_{SiC}$,
are ratioed to values obtained with magnesite at Epoch~1; $c_{SiC}$ is
ratioed its magnesite value at Epoch~2. The melilite fits plotted in
~\citetalias{Bowey2021} are slightly different because those were
ratioed to values obtained with melilite.

Fitted optical-depths of the bSiC bands were $\sim$ 10--50 per cent
higher than those obtained with foAk$_{91}$ and the increase in depth
was consistent between Epochs. The inferred rate of increase between
Epochs~2 and~5 is $1.1\pm 0.4$ per year.

\subsubsection{Minor changes in $c_{pah}$ and $c_{mag}$}
Values for $c_{pah}$  determined with magnesite were slightly higher at
Epochs 4B and 5 giving an average rate of ratioed $c_{pah}$ values 9
per cent higher ($0.48\pm0.01$ yr$^{-1}$); the higher rate of
$c_{pah}$ increase during Epochs 2--4B was 8 percent lower ($0.91 \pm
0.07$~yr$^{-1}$) than with foAk$_{91}$ due to the lower
$c_{pah}$-value in Epoch 4B. $c_{mag}$ is slightly lower than was
$c_{mel}$; the rate of increase in $c_{mag}$ is decreased by 20 per
cent ($0.32 \pm 0.04$ yr$^{-1}$) Within the uncertainties these values
remain consistent with \citetalias{Evans2020}'s $0.6 \pm
0.2$~yr$^{-1}$ estimate of the rate of increase in dust mass from the
emitted flux ($\lambda F_\lambda$). Absolute values of $\dot{c}_{pah}$, $\dot{c}_{pah}$(Ep2--4B), $\dot{c}_{SiC}$ (Ep2--5) and $\dot{c}_{mag}$ are
$0.038\pm0.01$, $0.073\pm0.06$, $0.02\pm0.01$, $0.011\pm0.001$, respectively, per year.
\subsubsection{Negligible changes in 8--13~\micron\ Fits}
\begin{figure}
\includegraphics[bb=280 143 540 751,width=0.8\linewidth,clip=]{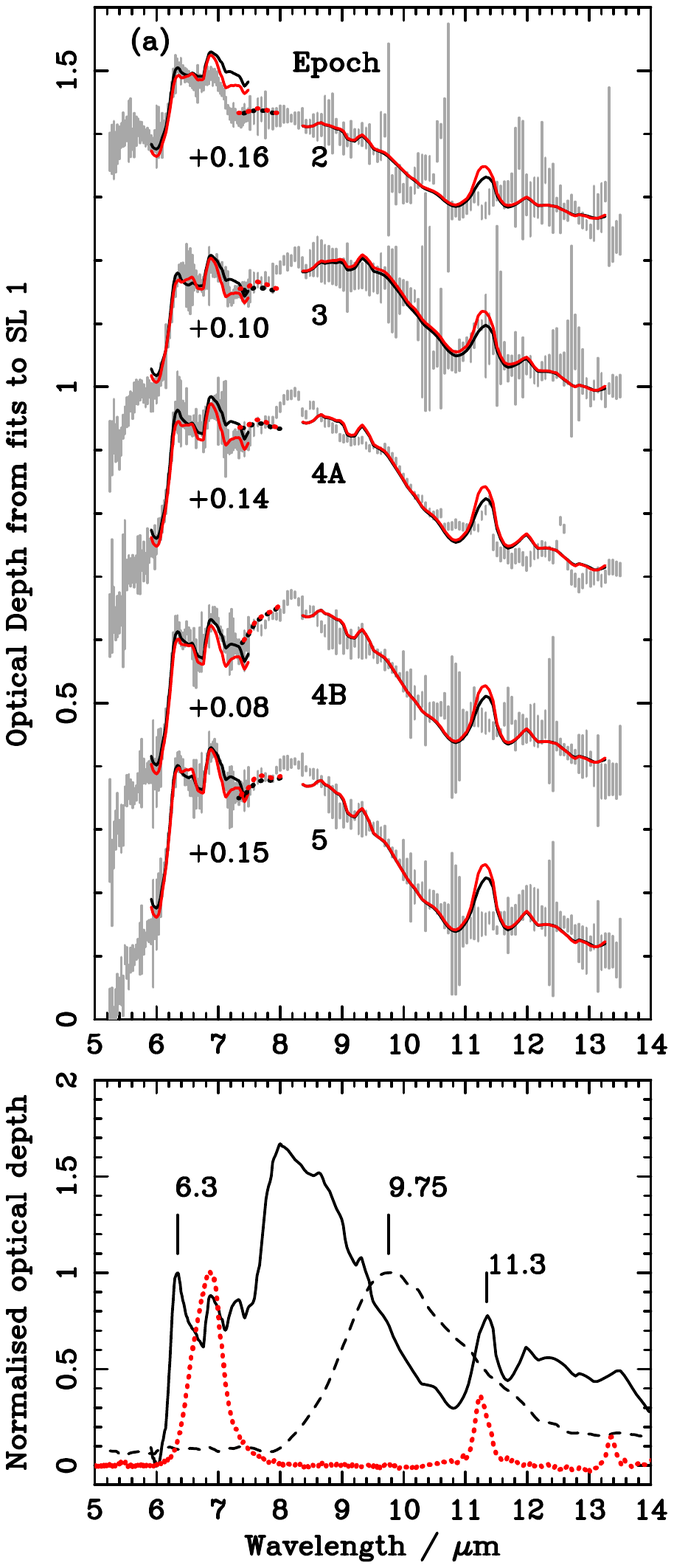}
\caption{Optical depth profiles (grey error bars) obtained with
  modelled 8.3--13.3~\micron~continua; offsets in the Y-axis are 0.0,
  0.3, 0.6, 0.9 and 1.2; 8--13~\micron\ solid curves are the fitted
  absorption features obtained with magnesite (red) and melilite
  (black), 6-7.5~\micron\ are the fitted features from the SL~2 offset
  in the Y-axis to match the level of the longer-wavelength fits
  (additional offsets are indicated); dotted curves between 7.5 and
  8.0~\micron\ indicate the contribution from bSiC extrapolated from
  the magnesite (red) and melilite (black) SL~2 fits. Magnesite (red
  dotted) with the other absorption profiles used to model
  8.3--13.3~\micron\ spectra: PAH spectrum (solid) and silicate
  absorption (dashed) modelled with the profile of interstellar dust
  towards Cyg OB2 no. 12. Modelling profiles are normalised at 6.3~
  and 9.75~\micron, respectively. Magnesite is normalised at
  6.9~\micron.\label{fig:sl1abs}}
\end{figure}
Fits of the 8.4--13.3~\micron\ spectra with revised values of
$c_{pah}$ and the original equation~\ref{eq:oldmod} component were
identical to those in \citetalias{Bowey2021}. New models
(equation~\ref{eq3}) including a magnesite component produced slightly
increased absorption at 11.2~\micron\ (Figure~\ref{fig:sl1abs}) which
increased the fitted value of $c_{sil}$ by 0--2 per cent, except in
Epoch 3 where the increase was 4 per cent. Once again the differences
are within the uncertainties even though magnesite (red) might
contribute to the 'PAH' feature at 11.3~\micron.

Differences in the shape of the modelled optically-thin emission were
negligible for the most optically-thin emission features. The
relatively unconstrained emission tempertures were 2--15~K warmer than
those in \citetalias{Bowey2021} in Epochs~1,~2,~4A, ~4B and~5; Epoch~3,
the least optically-thin and least constrained was nominally 82~K warmer.


\section{Sakurai's Object: Revised Column Densities and Mass Estimates}
\label{sec:massdensity}

\begin{table*}
\centering
\caption{Column Mass Density and Number Density for Dust Absorption Components}
\label{tab:codensity}
\begin{minipage}{\linewidth}
\begin{tabular}{ccccccccccccc}
\hline
&&\multicolumn{3}{c}{Mass Density, $\Sigma$}&\multicolumn{3}{c}{Number Density, $n$}&\multicolumn{3}{c}{Mass of 50~AU torus}&$\Sigma$&$n$\\
&&\multicolumn{3}{c}{$10^{-6}$~gcm$^{-2}$}&\multicolumn{3}{c}{cm$^{-2}$}&\multicolumn{3}{c}{$10^{-9}$M$_{\sun}$}&$10^{-6}$~gcm$^{-2}$&cm$^{-2}$\\
Dust && PAH    & bSiC  &magnesite& PAH    & bSiC  &magnesite&PAH&bSiC&magnesite&silicate&silicate\\
\multicolumn{2}{l}{Grain Size}&&&&53~nm&25~\micron&0.15~\micron&&&&&0.3~\micron\\

\hline
Epoch&Date\footnote{day-month-year}&&&&&&&\\

1  &15-04-2005&   $1.3_{0.67}^{6.1}$&54&1.2&2.2$\times 10^{10}$&1100   &$1.2\times 10^{8}$                       &1.1 &48&1.1  &--  &--\\
\\
2  &04-05-2007&   $2.0_{1.0}^{9.2}$&67 &2.1&3.4$\times 10^{10}$& 1300  &$2.1\times 10^{8}$  &1.8 &59&1.9  &27  & 3.0$\times 10^{8}$\\      
\\
3  &15-10-2007&   $2.5_{1.2}^{11}$&104 &2.5&4.3$\times 10^{10}$&2100   &$2.5\times 10^{8}$  &2.2 &92&2.2  &58  & 6.5$\times 10^{8}$\\
\\
4A &21-04-2008&  $3.1_{1.6}^{14}$&116 &2.4&5.3$\times 10^{10}$& 2300  &$2.4\times 10^{8}$  &2.7 &100&2.1 &54  & 6.1$\times 10^{8}$\\
\\
4B &30-04-2008&  $3.5_{1.7}^{16}$&123&2.2&6.0$\times 10^{10}$& 2500   &$2.2\times 10^{8}$  &3.1 &110&1.9 &46  & 5.2$\times 10^{8}$\\
\\
5  &18-10-2008&   $3.7_{1.8}^{17}$&167&2.8&6.4$\times 10^{10}$& 3300     &$2.8\times 10^{8}$  &3.3 &150&2.4 &42  & 4.7$\times 10^{8}$\\
\hline
WM\footnote{PAH, bSiC and melilite -from fits to weighted mean spectrum scaled to mean of Epochs 3--5 ($\uptau_{6.9}\simeq 0.24$); nSiC and astrosilicate from mean of Epochs 3--5.} &&$3.3_{1.6}^{14}$&128 &2.7&5.7$\times 10^{10}$& 2600  &2.7$\times 10^{8}$&&&&50& 5.6$\times 10^{8}$\\
\\
\multicolumn{3}{l}{Rates of increase / yr$^{-1}$}\\
Ep. 1--5\footnote{Italicised values for bSiC are based on Ep 2--5}&&0.64&{\it 73}&0.39&1.1$\times 10^{10}$&{\it 1500}&$0.39\times 10^8$&0.56&{\it 65}&0.34\\
Ep. 2--4B&&1.2   &&    &2.1$\times 10^{10}$  &&            &1.1&&\\
\hline
\end{tabular}
\end{minipage}
\end{table*}

Due to subtle changes in the fitted parameters and very significant
changes in the grain properties of the carrier of the broad
6.9-\micron\ band (see Table~\ref{tab:lab}) revised mass density,
number density, and dust-mass estimates are required: magnesite grain
sizes, absorption coefficients and indiviual grain masses are
respectively, 1.3 per cent, 12000~per~cent and $1.9\times 10
^{-4}$~per~cent of those attributed to foAk$_{91}$. In contrast with
the oxygen-rich melilites, which \citetalias{Bowey2021} assumed to be
formed in the external PN, we suggest the carbonaceous magnesite is
more likely to be be located in the 50~AU compact torus observed by
\citet{Chesneau2009} and gain an estimate of its mass.

\subsection{Mass and Column Densities}
As in \citetalias{Bowey2021}, the mass column density of absorbers,
$\Sigma$ along the line of sight is given by:
\begin{equation}
  \Sigma=\rho L= \uptau_{pk}/\kappa_{pk}.
\label{eq5}
\end{equation}
and the column number density of grains, $n$, is obtained by dividing
$\Sigma$, by the single-grain masses, $m_g$, in Table~\ref{tab:lab}.
We assume that the cold PAHs and bSiC grains are located in a
cylindrical volume of radius, $R=50$~AU and height of 50~AU and that
Sakurai's object is at a distance of 3.5kpc so that
\begin{equation}
M_{50 AU}\approx 8.8\times 10^{-4}\Sigma \ M_{\sun}
\end{equation}
and the rate of mass increase can be deduced by replacing $\Sigma$
with $\dot{\Sigma}$.

Rates of mass increase are determined by multiplying the gradients of
the increases by the fitted constants $c_{pah}$, $c_{mag}$ at Epoch~1,
or Epoch~2 ($c_{SiC}$) in Figure~\ref{fig:dmi} as
appropriate\footnote{\citetalias{Bowey2021} did not remember to
multiply the gradients by Table~\ref{tab:sl2fits} at Epoch~1 before
quoting them, to correct this use the values for Epoch~1 in her
Table~3, namely $c_{pah}=0.085$ and $c_{mel}=0.031$}.

Values of $\Sigma$, $n$ and torus-mass for each carbonaceous dust
component and are listed in Table~\ref{tab:codensity}: values for
PAHs and bSiC are revised, the magnesite components are new, and the
foreground silicate column densities are unchanged. Rates of increase,
$\dot{\Sigma}$, $\dot{n}$, and $\dot{M}_{50AU}$ in the carbonaceous
components are listed below the main table.

\subsection{Rates of increase}
\label{sec:increaserate}
The rate of increase in PAH mass-density,
$\dot{\Sigma}_{pah}=0.64\times10^{-6}$gcm$^{-2}$ in Epochs 1--5, is 3
per cent higher with magnesite than with foAk$_{91}$; with magnesite
the enhanced rate between Epochs~2--4B is twice the average rate for
Epochs 1--5 at $1.2\times10^{-6}$gcm$^{-2}$yr$^{-1}$, but 14 per cent
lower than obtained with foAk$_{91}$. This gives $n_{pah}$ increase
rates of 1.1 to 2.1 $\times10^{10}$ grains cm$^{-2}$yr$^{-1}$ and an
increase in the torus mass of
$0.56-1.1\times10^{-9}$M$_{\sun}$yr$^{-1}$. The enhanced rate is 10
times lower than Evans estimate for May 1999 (11$\times 10^{-9}$)
based on a different set of assumptions but the increased rate
probably still correlates with the period of enhanced mass loss
identified by \citet{Evans2020} and \citet{Tyne2002} who estimated a
3.6 ratio in mass-loss rate between May 1999 and September 2001.


For bSiC, $\dot{\Sigma}_{SiC} =73\times10^{-6}$gcm$^{-2}$yr$^{-1}$ or
1500 grains cm$^{-2}$yr$^{-1}$; the mass increase rate is 65$\times
10^{-9}$~M$_{\sun}$ yr$^{-1}$ between Epochs 2 and 5. For magnesite,
the mean rate of mass density increase is
$\dot{\Sigma}_{mag}=0.39\times10^{-6}$gcm$^{-2}$yr$^{-1}$ or
$3.9\times 10^7$ grains cm$^{-2}$yr$^{-1}$; if carbonates are located
in the torus the mass increase rate is 0.34$\times 10^{-9}$~M$_{\sun}$
yr$^{-1}$.

\subsection{Results}
These models replace fits with large melilite grains in
\citetalias{Bowey2021} for fits with submicron-sized carbonate grains.
The modelled PAH column densities and torus masses for each Epoch are
the same to within $\pm10$ per~cent; the fitted component of large SiC
grains (bSiC) is better constrained and 10--50~per~cent
higher. Astrosilicate column densities are unchanged.  These results
are not sufficiently altered to affect \citetalias{Bowey2021}'s
conclusions about the formation time of the PAHs or bSiC, but the link
between astrosilicate coagulation to melilites is removed as is the
connection between these observations and the abundance of melilite in
meteorites.

Due to differences in formation rate, carbonate (magnesite) mass
densities and torus masses are similar to the those inferred for PAHs
for Epochs 1--3, but lower for Epochs 4A--5. Ratios of particle number
density for silicates and magnesite vary between 1.4 (Epoch~2) to 2.5
(Epochs 3-4B) to 1.7 (Epoch 5). The ratio of PAH to magnesite
particles is 170 (Epochs 1--3), to 230 (Epoch 4A-5). The ratio of PAH
to SiC number density is $\sim 2.3 \times 10^7$.

\section{How are astronomical carbonates formed?}
\label{sec:carbform}
During the period of the observations, the column density of magnesite
in Sakurai's Object increased at a rate of $0.39\times
10^{-6}$gcm$^{-2}$ yr$^{-1}$, about 0.6 of the average rate of
increase in PAH mass (Section~\ref{sec:increaserate}). Therefore it
seems that carbonates are either condensing directly from gas, or
other dust components are being converted to carbonates. The
experiments discussed below suggest carbonates condense from gas below
700~K and decompose above about 730~K. The inferred carbonates in Sakurai's
Object appear as absorption features in dust which is cooler than the
280--230~K continuum used in modelling the absorption features (see
models in Section~\ref{sec:sl2fits}).

\subsection{Direct condensation from the gas is problematic}
Direct condensation from the gas is problematic because chemical
equilibrium and non-equilibrium models suggest <1 per cent of the dust
condensed from gases outflowing from oxygen-rich AGB stars could be
CaCO$_3$ \citep[or another carbonate][]{FG2005} and that the amount
formed in a carbon-rich envelope like that of Sakurai's Object will be
$\simeq0$ due to the absence of free oxygen. Analogous laboratory
experiments might be chemical vapour deposition in a CO-rich or
CO$_2$-rich and oxygen- and H$_2$O-poor atmosphere. We have found none
for a CO atmosphere. The nearest we can find are:
\begin{enumerate}
\item an experiment by
\citet{Sulimai2021} who formed thin films of calcium carbonate on FTO
(fluorine-doped tin) glass substrates by chemical vapour deposition
(CVD) from heated CaCl$_2$ solution and CO$_2$ gas at a flow rate of
100 standard cubic centimetres per minute at 1 bar. CaCO$_3$ formed at
temperatures of 623--723~K with a peak formation rate at 673~K.

\item Unidentified carbonate contamination in the infrared spectra of
  vapour-condensed CaO and Ca(OH)$_2$ smokes published
  by~\citet{KN2005}. In these experiments Ca metal was placed in a
  graphite boat inside an alumina furnace tube and exposed to a
  hydrogen-rich atmosphere at 0.1 bar and temperatures between 1080
  and 1170~K. Spectra of their run products are similar to our
  carbonate spectra with characteristic narrow peaks near 11.4 and
  14.0~\micron\ as well as the band at 6.9~\micron. We conclude that
  the disordered sample picked up CO$_2$ during the experiment or from
  the air\footnote{The infrared carbonate bands are known to appear
  rapidly when CaO is obtained by burning limestone (CaCO$_3$) in air
  \citep{GR2009}} as well as the discussed water to form a surface
  layer of carbonate before their spectra were obtained. The main
  31-\micron\ peak in CaO has a shoulder at
  17.9~\micron\ \citep{HKS2003} which appears in the \citet{KN2005}
  data which extend to 25~\micron.
\end{enumerate}


\subsection{Formation from older dust}
Does the small reduction in the optical depth of the foreground
silicate feature indicate that silicates are being converted to
carbonates? \citet{Day2013} have formed metastable calcium carbonate
(vaterite and calcite) by exposing room-temperature amorphous
CaSiO$_3$ glass in a sapphire high-pressure tube to gaseous CO$_2$ at
a pressure of 6~bar - well above ambient astronomical pressures. Their
carbonate is stable at temperatures below 753~K; above this it starts
to anneal and break down into CaSiO$_3$ and
CO$_2$. \citet{Sulimai2021} also showed that CaCO$_3$ decomposes
between 723~K and 873~K. If carbonates also fit the broad
6.9-\micron\ band in YSOs and molecular clouds it seems likely that
exposure of pre-existing grains to CO$_2$ and water might also result
in carbonate formation but appropriate experiments are required to
prove this.



\section{Summary}
\label{sec:conclusion}
We present new 600--2000cm$^{-1}$ room-temperature spectra of 3
carbonates: calcite, dolomite and magnesite which have strong broad
peaks centred at 6.97, 6.87 and 6.87~\micron, respectively. These
peaks resemble an astronomical absorption feature observed in a
variety of dense astronomical environments including young stellar
objects and molecular clouds. We focus on fitting spectra of low-mass
post-AGB star Sakurai's Object which has been forming substantial
quantities of carbonaceous dust since an eruptive event in the 1990s.

Comparison of published melilite-group and hibonite overtone spectra
obtained by \citet{BH2005} with our carbonate spectra shows that they
are dominated by features due to previously unrecognised carbonate
contamination. Thus \citet{BH2005} and \citetalias{Bowey2021} were
fitting the astronomical data with bands due to carbonates.

New overtone spectra obtained from uncontaminated melilite samples
contain two $\sim 0.1$~\micron-wide peaks centred at 6.4 and
6.8~\micron; the mass absorption coefficient of the 6.8~\micron~peak
is $\lesssim 0.05$~per~cent of the value for carbonates. These bands
are too narrow and too weak to provide a measurable contribution to
the feature in Sakurai's Object and would probably not contribute to
other features either. This removes the proposed links between melilites in
Sakurai's Object and the abundance of melilite in meteorites.

Replacement of melilite with magnesite or dolomite in models of
Sakurai's Object preduces statistically better fits to the
6--7.5-\micron\ Spitzer spectra without significantly altering other
conclusions of Bowey's previous models. The main difference is a
10--50~per~cent-larger component due to 25~\micron-sized SiC grains
which increases consistently between epochs. If the carbonates
contribute to the mass of the 50~AU torus, the mass is similar to the
mass of PAH dust between 2005 April and 2007 October, but lower
between 2008 April and 2008 October. On average PAH particles are
$2.3\times 10^7$ more abundant than SiC grains and 200 times more
abundant than magnesite grains, based on their respective particle
sizes of 53nm, 25~\micron\ and 0.15~\micron.

Experimental evidence relevant to carbonate formation in low pressure
astronomical environments is scarce, although carbonates appear to be
stable at temperatures below about 700~K. Evaluating a feasible
mechanism is thus challenging. Direct condensation from outflowing gas
has been invoked for many species. Alternatively, silicates are being
converted to carbonates or they are being made by other chemical
changes in pre-existing dust. The small reduction in the optical depth
of the foreground silicate feature might indicate that silicates are
being converted to carbonates.

\section*{Acknowledgements}
We thank the reviewer, Joseph Nuth, for a positive and helpful report.
JEB and the purchase of samples for the experiments were funded by a
2-yr Science and Technology Research Council Ernest Rutherford
Returner Fellowship (ST/S004106/1) plus a 6 month extension from
Cardiff University. AMH was supported by the USA National Science
Foundation. We thank Paul Carpenter of Washington University for the
microprobe analyses. LR Spitzer spectra were obtained from the Combined
Atlas of Sources with Spitzer/IRS Spectra (CASSIS), a product of the
Infrared Science Center at Cornell University, supported by NASA and
JPL. Observations were made with the Spitzer Space Telescope, which
was operated by the Jet Propulsion Laboratory, California Institute of
Technology under a contract with NASA.

\section*{Data Availability}

The carbonate and true melilite overtone spectra presented in this
article are subject to a partial embargo of 12 months from the
publication date of the article during which the data will be
available from the authors by request. Once the embargo expires the
data will be available from
https://zenodo.org/communities/mineralspectra/






\appendix
\section{Correction of Ak$_{70}$ and Ak$_{13}$ spectra}
\label{app:base}
\begin{figure}
\includegraphics[bb=10 0 520 730,width=\linewidth,clip=]{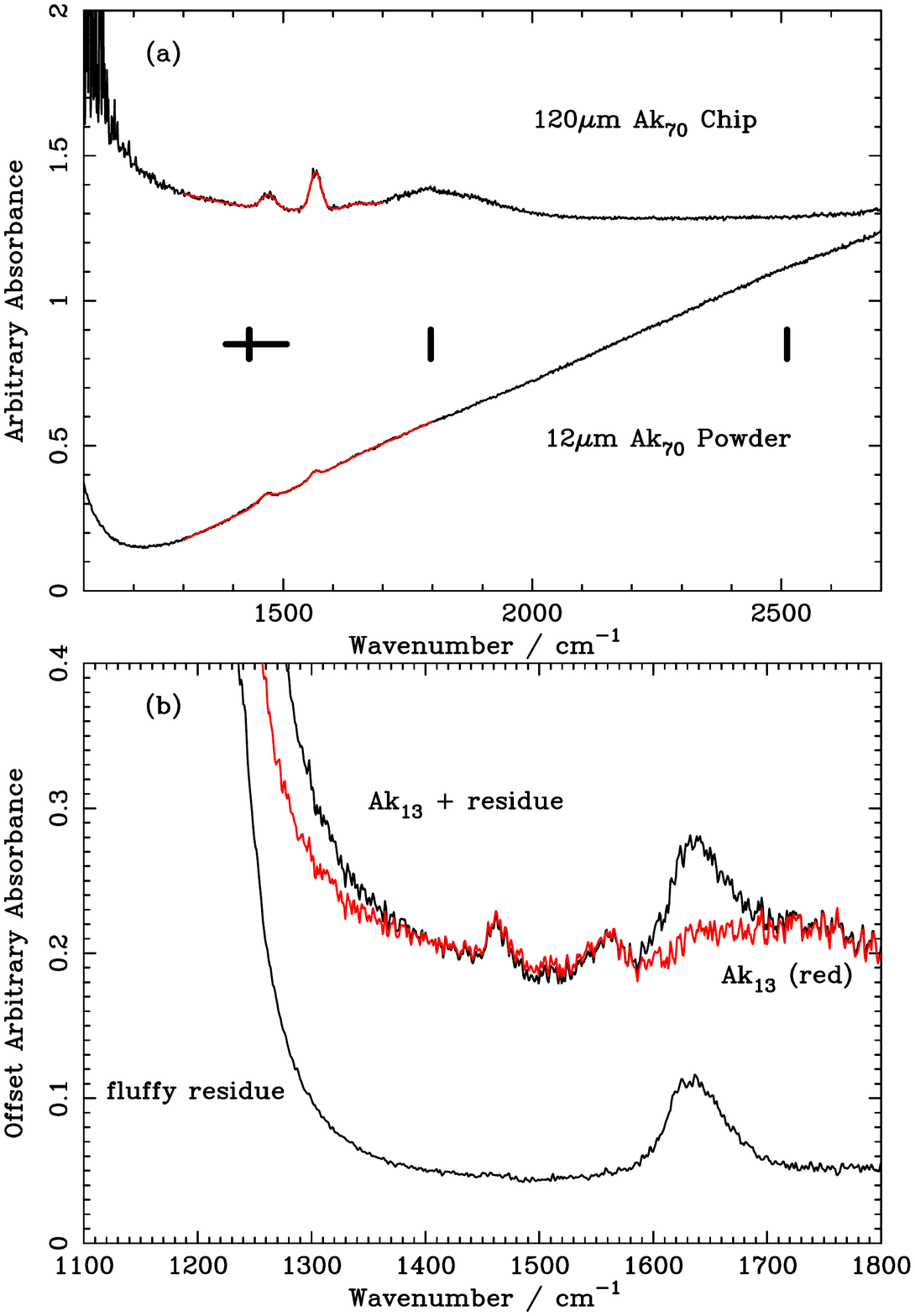}
\caption{(a) Comparison between spectra of a melilite Ak$_{70}$ chip and
  the powdered mineral before baseline removal. Spectra are shown with
  the fitted baselines and Gaussian peaks (red). Vertical bars
  indicate the likely positions of narrow bands due to carbonate
  contamination, the crossed bar indicates the FWHM of the broadest
  carbonate peak centred near 1432~cm$^{-1}$. There is no carbonate
  contamination in this sample. (b) Subtraction of feature due to the unidentified fluffy residue with the Ak$_{13}$ crystals.}
\label{fig:bbase}
\end{figure}

Local baselines were subtracted from the spectra of the Ak$_{70}$
chips and powder film (Figure~\ref{fig:bbase}(a)) by fitting three
peaks at 1466, 1563 and 1647~cm$^{-1}$. The chip spectra also
exhibited a broad peak at 1800~cm$^{-1}$ which might be due to H$_2$O
inclusions in the crystal. There is no hint of carbonate contamination
in this spectrum or in Ak$_{13}$ because the broad 1432~cm$^{-1}$ peak
and the narrow peaks are absent.

The Ak$_{13}$ spectrum was obtained from crystals embedded in an
unknown fluffy residue with a peak at $\sim 1630$cm$^{-1}$
(Figure~\ref{fig:bbase} (b)). Baselines were subtracted from the
contaminated spectrum and the fluff spectrum. The fluff spectrum was
removed by subtracting its scaled peak from the contaminated Ak$_{13}$
spectrum. The result (red) was then fitted with a 2nd order polynomial
and three Gaussian peaks between 1340 and 1750 cm~$^{-1}$. The
polynomial was subtracted to produce the spectrum in
Figure~\ref{fig:trumel}.

\bsp	
\label{lastpage}
\end{document}